\documentstyle[aps,epsfig]{revtex}                  
\begin{document}
\pagestyle{empty}                                      
\preprint{ \font\fortssbx=cmssbx10 scaled \magstep2 \hbox to
\hsize{ \hfill$\raise .5cm\vtop{\hbox{NCTU-HEP-0202}}$}}
\draft
\vfill
\title{Strong-Field QED and the Inverse Mellin Transform\footnote{Talk presented in Fifth Workshop on Quantum
Field Theory under the Influence of External Conditions, Leipzig,
Germany, 10-14 Sep 2001. } }

\author{ Guey-Lin Lin } %

\address{Institute of Physics, National Chiao Tung University,
Hsinchu 300, Taiwan}
\date{\today}
%
%
\vfill
\maketitle
\begin{abstract}
We introduce the technique of inverse Mellin transform in a problem
of strong-field QED.  We show that the {\it moments} of pair
production width in a uniform background magnetic field are
proportional to the derivatives of photon polarization function at
the zero momentum. Hence, the pair-production width or the
absorptive part of the photon polarization function is
calculable from the latter by the inverse Mellin transform. Using
the {\it Kramers-Kronig} relation, the dispersive part of photon
polarization function can be computed as well. Therefore the
analytic property of the photon polarization function in all energy
range is obtained. We also discuss briefly the possible extensions of this
technique to other problems.
\end{abstract}
%
%
%
%
\pagestyle{plain}

\section{Introduction}

The wave function and energy
quantization of a relativistic charged particle in a uniform
background magnetic field is well understood. For a
relativistic electron (positron), one solves for the Dirac
equation in a background magnetic field. The solution is
characterized by the energy levels:
\begin{equation}
E^2_{n,s_z}=m_e^2+p_z^2+eB\left(2n+1+2s_z\right),
\end{equation}
with $s_z$  the electron spin projection along the $+z$
direction, and the correspondent wave function
\begin{equation}
\psi^{\pm}_{n,p_y,p_z,s_z}(\vec{r},t)= \exp(-iE_n t+ip_y y+ip_z
z){\bf F}^{\pm}_{n,p_z,s_z}(x'),
\end{equation}
where ${\bf F}^{\pm}_{n,p_z,s_z}(x')$ is a four-component spinor
with $x'=x-p_y/eB$. The detailed form for ${\bf
F}^{\pm}_{n,p_z,s_z}$ can be found, for example, in\cite{JL}.

In spite of our knowledge in the above energy quantization and
wave function, it is often non-trivial to compute a
physical process occurring in a background magnetic field,
particularly, if there are more than one charged fermion
involving in the reaction. A well-known example is the pair
production process, $\gamma\to e^+ e^-$, which is relevant to the
gamma-ray attenuation in the neutron star\cite{STU}, and was
studied first by Toll\cite{toll} and Klepikov\cite{kle} independently.
Both authors computed the pair production width by squaring the
$\gamma\to e^+ e^-$ matrix elements using exact electron and
positron wave functions in a background magnetic field, and summing
over all available final states consistent with the initial photon energy.
The most updated calculation using this approach was given in
\cite{DH}. We note that all of the above works assume $q^2=0$
for the photon momentum, despite the presence of background magnetic field.
To apply the precise photon dispersion relation in the
calculation of pair-production width, one needs to study the
photon polarization function in a background magnetic field. In fact, a
consistent calculation of pair-production width require the
knowledge of both absorptive and dispersive parts of photon
polarization function. In
this regard, Tsai and Erber\cite{TE} obtained the absorptive part
of the one-loop photon polarization function in the asymptotic
limit $\omega\gg 2m_e$ and $B\ll B_c\equiv m_e^2/e$ with the
assumption $q^2=0$. Their result was shown\cite{TE} to agree with
that of Toll and Klepikov. However, in the above
asymptotic limit, the threshold behavior of the pair-production
width is completely lost. Later on, Shabad\cite{shabad} obtained
the absorptive part (and the dispersive part as well) of one-loop
photon polarization function for a general photon energy and
magnetic-field strength. In that work, the threshold behavior of
the pair-production width was worked out explicitly. We remark
that Refs.\cite{TE,shabad} employed Schwinger's proper-time
representation for the electron and positron Green's
functions\cite{sch} which make up the photon polarization function. To
our knowledge, Ref.\cite{shabad} is the first work demonstrating
that the proper-time representation for the photon polarization
function gives equivalent pair-production width to that given by
squaring the $\gamma\to e^+ e^-$ amplitude directly.
Unfortunately, the manipulations of Ref.\cite{shabad} are rather
involved and the substantial details of them were given in some
other unpublished preprints\cite{prep}. It is not very clear how
one can generalize the approach of Ref.\cite{shabad} to other
processes.

In this report, we provide an alternative derivation of the
pair-production width (or equivalently the absorptive part of the
photon polarization function) from the proper-time representation
of the photon polarization function. We shall also outline the
procedure of obtaining the dispersive part of the polarization
function, which is required for computing the photon index of refraction. It
will be clear that our approach is very straightforward and
physically intuitive. This report is organized as follows: In
Section 2, we derive the sum rule that relates the moments of the
pair-production width to the derivatives of the photon
polarization function at the zero momentum. In Section 3, we apply
this sum rule to the photon polarization function in an asymptotic
limit, i.e., $\omega\gg 2m_e$ and $B\ll B_c$. A general analysis
valid for arbitrary $\omega$ and $B$ is given in Section 4. We
conclude in Section 5.

\section{The Sum Rule}

We are interested in the properties of photon
polarization function in the background magnetic field,
$\Pi_{\mu\nu}$. Since a photon has two different polarization
states in a background magnetic field, one may choose the
polarization states as $\epsilon^{\mu}_{\parallel}\equiv (0,\vec{
\epsilon}_{\parallel})$ and $\epsilon^{\mu}_{\bot}\equiv (0, \vec{
\epsilon}_{\bot})$ where $\vec{\epsilon}_{\parallel}$ is lying on
the plane spanned by the photon momentum ${\bf q}$ and the
magnetic field vector ${\bf B}$, while $\vec{\epsilon}_{\bot}$ is
perpendicular to the plane. It is then convenient to define the
scalar functions
$\Pi_{\parallel,\bot}=\epsilon^{\mu}_{\parallel,\bot}\Pi_{\mu\nu}
\epsilon^{\nu}_{\parallel,\bot}$ which govern the behaviors of
polarization states $\epsilon^{\mu}_{\parallel,\bot}$ in the
background magnetic field. The sum rule for
$\Pi_{\parallel,\bot}$ is derived by considering the following
contour integral\cite{KLT1}:
\begin{equation}
I_n=\int_C {d\omega^2\over 2\pi i}{\Pi_{\parallel,\bot}(\omega^2)
\over (\omega^2+\omega_0^2)^{n+1}},
\end{equation}
where the integration contour $C$ is shown in Fig. 1. We have
implicitly assume $q^2=0$ in the above equation in order to
compare with previous results\cite{TE}. This assumption will be
relaxed later. For convenience, let us take ${\bf B}$ to be along
$+z$ direction, while the photon momentum ${\bf q}$ is on the $XZ$
plane making an angle $\theta$ to ${\bf B}$. Hence $q$ can be
parameterized as $q^{\mu}=(\omega, \omega \sin\theta, 0, \omega
\cos\theta)$. We note that the integral $I_n$ may be evaluated in
two different ways. One computes $I_n$ either by the residue
theorem or by a direct integration along the contour $C$ with the
realization that the contribution from the outer circle vanishes.
The equivalence of two integration procedures gives rise to the
relation:
\begin{equation}
{1\over n!}\left({d^n\over d(\omega^2)^n}
\Pi_{\parallel,\bot}\right)\Big{\vert}_{\omega^2=-\omega_0^2}
={1\over \pi}\int_{M^2_{\parallel,\bot}}^{\infty}d\omega^2 {{\rm
Im}\Pi_{\parallel,\bot}(\omega^2)\over
(\omega^2+\omega_0^2)^{n+1}}, \label{dispersion}
\end{equation}
where $M_{\parallel,\bot}$ are the threshold energies for pair
productions, given by
\begin{equation}
M^2_{\parallel}\sin^2\theta=4m_e^2, \;
M^2_{\bot}\sin^2\theta=m_e^2\left(1+\sqrt{1+2{B\over B_c}}
\right)^2.
\end{equation}
Since the absorption coefficient (pair production width) is given by
$\kappa_{\parallel,\bot}={\rm
Im}\Pi_{\parallel,\bot}/\omega$, Eq.~(\ref{dispersion})
relates the real part of photon polarization function to
the absorption coefficient. One can further simplify
Eq.~(\ref{dispersion}) by taking $\omega_0^2=0$ such that
\begin{equation}
{1\over n!}\left({d^n\over d(\omega^2)^n}
\Pi_{\parallel,\bot}\right)\Big{\vert}_{\omega^2=0}
={M_{\parallel,\bot}^{1-2n}\over
\pi}\int_{0}^{1}dy_{\parallel,\bot}\cdot y_{\parallel,\bot}^{n-1}
\cdot \left(\kappa_{\parallel ,\bot}
y_{\parallel,\bot}^{-1/2}\right), \label{mellin}
\end{equation}
with $y_{\parallel,\bot}=M_{\parallel,\bot}^2/\omega^2$. One notes
that the absorptive part of $\Pi_{\parallel,\bot}(\omega^2)$
vanishes for the range $0 \le \omega^2 \le
M^2_{\parallel,\bot}$\cite{adl}. Therefore one can effectively set
the integration range of Eq.~(\ref{mellin}) as from
$y_{\parallel,\bot}=0$ to $y_{\parallel,\bot}=\infty$. As
mentioned before, the above sum rule is derived by assuming
$q^2=0$. For a general $q^2$, the sum rule becomes\cite{KLT2}:
\begin{equation}
{1\over n!}\left({d^n\over d(q_{\parallel}^2)^n}
\Pi_{\parallel,\bot}\right)\Big{\vert}_{q_{\parallel}^2=0}
={m_{\parallel,\bot}^{-2n}\over
\pi}\int_{0}^{\infty}du_{\parallel,\bot}\cdot
u_{\parallel,\bot}^{n-1} \cdot \left(\omega \kappa_{\parallel
,\bot}\right),\label{mellin2}
\end{equation}
where $m_{\parallel}^2=4m_e^2$, $m_{\bot}^2=m_e^2(1+\sqrt
{1+2B/B_c})^2$, $q_{\parallel}^2=\omega^2-q_z^2$, and
$u_{\parallel,\bot}=m_{\parallel,\bot}^2/q_{\parallel}^2$. Both of
Eqs.~(\ref{mellin}) and (\ref{mellin2}) indicate that the {\it
moments} of absorption coefficients are proportional to the
derivatives of photon polarization functions at the zero value of
$q_{\parallel}^2\equiv \omega^2-q_z^2$. Hence the absorption
coefficients can be calculated from the derivatives of photon
polarization functions at the zero value of $q_{\parallel}^2$ by
the inverse Mellin transform. \vspace{3cm}
\begin{figure}
  \unitlength 1mm
   \begin{center}
      \begin{picture}(25,100)
     \put(-70,30) {\epsfig{file=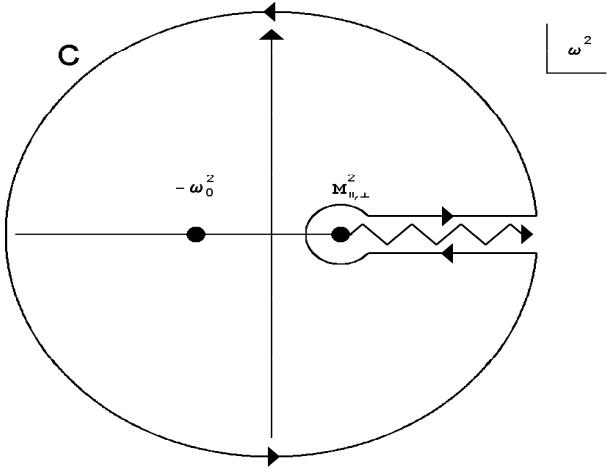,width=10cm,height=10cm}}
      \end{picture}
   \end{center}
\vspace{-3cm}
\caption{The integration contour for $I_n$ and the
analytic structure of $\Pi_{\parallel,\bot}$. In actual
calculations, we take the radius of the circle to infinity.}
\label{fig1}
\end{figure}

\section{The behaviors of $\Pi_{\parallel,\bot}$
in the asymptotic limit}

To study the behaviors of
$\Pi_{\parallel,\bot}$, let us begin with the proper-time
representation of photon polarization function $\Pi_{\mu\nu}$ in a
uniform background magnetic field\cite{TS}:
\begin{eqnarray}
\Pi_{\mu\nu}(q)&=&-{e^3B\over (4\pi)^2}\int_0^{\infty}ds
\int_{-1}^{+1} dv \{e^{-is\phi_0}[(q^2g_{\mu\nu}-
q_{\mu}q_{\nu})N_0 \nonumber \\
&-&(q_{\parallel}^2g_{\parallel\mu\nu}-
q_{\parallel\mu}q_{\parallel\nu})N_{\parallel}
+(q_{\bot}^2g_{\bot\mu\nu}-
q_{\bot\mu}q_{\bot\nu})N_{\bot}]\nonumber \\
&-&e^{-ism_e^2}(1-v^2)(q^2g_{\mu\nu}- q_{\mu}q_{\nu})\},
\label{proper_t}
\end{eqnarray}
where the photon momentum has been decomposed into
$q_{\parallel}^{\mu}\equiv (\omega, 0, 0, q_z)$ and
$q_{\bot}^{\mu}\equiv (0, q_x, q_y, 0)$. The phase $\phi_0$ and
the functions $N_0$, $N_{\parallel}$, and $N_{\bot}$ are given by
\begin{equation}
\phi_0=m_e^2-{1-v^2\over 4} q_{\parallel}^2-{\cos(zv)-\cos(z)\over
2z\sin(z)} q_{\bot}^2
\end{equation}
with $z=eBs$, and
\begin{eqnarray}
N_0&=&{\cos(zv)-v\cot(z)\sin(zv)\over \sin(z)},\nonumber \\
N_{\parallel}&=&-\cot(z)\left(1-v^2+{v\sin(zv)\over
\sin(z)}\right)
+{\cos(zv) \over \sin(z)},\nonumber \\
N_{\bot}&=&-{\cos(zv)\over \sin(z)} +{v\cot(z)\sin(zv)\over
\sin(z)}+ 2{\cos(zv)-\cos(z)\over \sin^3(z)}.
\end{eqnarray}
In the limit $B\ll B_c$ with $q^2=0$, we write the derivative of
$\Pi_{\parallel,\bot}$ into an asymptotic series in $B/B_c$:
\begin{eqnarray}
{1\over n!}\left({d^n\over d(\omega^2)^n}
\Pi_{\parallel,\bot}\right)\Big{\vert}_{\omega^2=0}&=&{2\alpha_e
m_e^2\over \pi}\left({B^2\sin^2\theta\over 3B_c^2m_e^2}\right)^n
{\Gamma(3n-1)\Gamma^2(2n)\over \Gamma(n) \Gamma(4n)}\nonumber \\
&\times&\left({6n+1, 3n+1\over 4n+1}\right)+\cdots, \label{diff}
\end{eqnarray}
where the disregarded terms are of higher order in $B/B_c$. From
the sum rule in Eq.~(\ref{mellin}), we obtain the absorption
coefficients $\kappa_{\parallel,\bot}$ via the inverse Mellin
transform\cite{KLT1}:
\begin{eqnarray}
\kappa_{\parallel}(\lambda^{'})&=&{\alpha m_e^2\over i\pi\omega}
\int_{-i\infty+a}^{+i\infty+a} ds {(\lambda^{'})}^{2s}
{\Gamma(3s)\Gamma^2(2s)\over \Gamma(s)\Gamma(4s)}{1\over 3s-1}
\times {6s+1\over 4s+1},\nonumber \\
\kappa_{\bot}(\lambda^{''})&=&{\alpha m_e^2\over
i\pi\omega}{2\over 1+\sqrt{1+2B/B_c}}
\int_{-i\infty+a}^{+i\infty+a} ds {(\lambda^{''})}^{2s}
{\Gamma(3s)\Gamma^2(2s)\over \Gamma(s)\Gamma(4s)}{1\over 3s-1}
\times {3s+1\over 4s+1},\nonumber \\
\label{abso1}
\end{eqnarray}
where $a$ is any real number greater than $1/3$; while
$\lambda^{'}= (\omega\sin\theta B/ \sqrt{3}m_e B_c)$ and
$\lambda^{''}=\lambda^{'}\cdot (1+\sqrt{1+2B/B_c})/2$. These
absorption coefficients were shown\cite{KLT1} to
agree with previous results by Tsai and Erber\cite{TE} who
obtained $\kappa_{\parallel,\bot}={\alpha\over
2}\sin\theta\left(eB/ m_e\right)
T_{\parallel,\bot}(\lambda)$ with $\lambda={3\over
2}(eB/m_e^2)(\omega/m_e)\sin\theta$, and
\begin{equation}
T_{\parallel,\bot}(\lambda)={4\sqrt{3}\over \pi \lambda} \int_0^1
dv (1-v^2)^{-1}\left[(1-{1\over 3}v^2), ({1\over 2}+{1\over 6}v^2)
\right]K_{2/3}\left({4\over \lambda}{1\over 1-v^2}\right),
\label{bessel}
\end{equation}
with $K_{2/3}$  the modified Bessel function.

It is important to point out that both of $\kappa_{\parallel}$ and
$\kappa_{\bot}$ are smooth functions of the photon
energy. This behavior actually does not describe the
real physical situation where new absorption peaks keep on
emerging as the photon energy increases. One attributes this
problem to the asymptotic expansion performed in Eq.~(\ref{diff}). As
it is easily seen that, apart from the trivial mass factor
$m_e^{2-2n}$, the first term on the R.H.S. of Eq.~(\ref{diff})
grows to infinity as $n$ increases, no matter how small the ratio
$B/B_c$ is. This then implies that the disregarded higher-order terms
are in fact non-negligible for a sufficiently large $n$. As a
result, the large moments of $\kappa_{\parallel,\bot}$ are not
accurately determined by the first term of the above asymptotic
expansion. Therefore, to obtain a correct energy dependencies of the
absorption coefficients $\kappa_{\parallel,\bot}$, one should not
expand in $B/B_c$ even for a small magnetic-field strength.

\section{The exact one-loop result for $\Pi_{\parallel,\bot}$}

In this section, we calculate $\Pi_{\parallel,\bot}$ for
a general photon energy and background magnetic-field strength.
From Eq.~(\ref{proper_t}), we obtain the integral representation
for the scalar functions $\Pi_{\parallel,\bot}$, i.e.,
$\Pi_{\parallel}(\omega^2,{\bf q}_{\bot}^2,\theta)
=-(\alpha_e\omega^2 \sin^2\theta /4\pi)\bar{\Pi}_{\parallel}$ and
$\Pi_{\bot}(\omega^2,{\bf q}_{\bot}^2,\theta) =-(\alpha_e {\bf
q}_{\bot}^2/ 4\pi) \bar{\Pi}_{\bot}$, with
\begin{eqnarray}
\bar{\Pi}_{\parallel}&=& \int _0 ^\infty dz \int _{-1}^{+1} dv \;
\exp [ {-is \phi_0}]
N_{\parallel}, \nonumber \\
\bar{\Pi}_{\bot}&=& \int _0 ^\infty dz \int _{-1}^{+1} dv \; \exp [ {-is
\phi_0}] N_{\bot},
\end{eqnarray}
where ${\bf q}_{\bot}^2=q_x^2+q_y^2$, and $\theta$ is the angle
between the photon momentum ${\bf q}$ and the magnetic field
vector ${\bf B}$. The calculations of
$\bar{\Pi}_{\parallel}$ and $\bar{\Pi}_{\bot}$ are non-trivial. For illustrations,
we will only show the details of computing one particular term in
$\bar{\Pi}_{\parallel}$. Let us consider the following integral
\begin{equation}
\bar{\Pi}_{\parallel}^A= \int _0 ^\infty dz \int _{-1}^{+1} dv \;
\exp [ {-is \phi_0}]\frac{\cos(zv)}{\sin(z)},
\end{equation}
where $\cos(zv)/ \sin(z)$ is the last term in $N_{\parallel}$. To
apply the sum rule given by Eq.~(\ref{mellin2}), we rotate the
integration contour $s\to -is$ and write $\bar{\Pi}_{\parallel}^A$
in an infinite series of $q_{\parallel}^{ 2}$, i.e.,
\begin{eqnarray}\label{Aab}
\bar{\Pi}_{\parallel}^A(q_{\parallel}^{\prime 2},{\bf
q}_{\bot}^{\prime 2})&=&
 {\sum}K^A_{lmpp'}({\bf q}_{\bot}^{\prime 2})
    (\frac{1}{B^{\prime}})^n \frac{(q_{\parallel}^{\prime 2})^{n-l}}{(n-l)!}
\int _0 ^\infty dz \; z^{n-l} \; \exp [ -z\beta]\nonumber \\
&\times&\int _{0} ^{1} dv \; (1-v^2)^{n-l} \left(\exp [ \alpha
zv]+\exp [ -\alpha  zv]\right),
\end{eqnarray}
with ${\sum} \equiv \sum_{n=0}^{\infty} \sum_{l=0}^n
\sum_{m=0}^{\infty}\sum_{p',p=0}^l$, $q_{\parallel}^{\prime 2}=
q_{\parallel}^2/4m_e^2$, $B'= B/B_c$, ${\bf q}_{\bot}^{\prime 2}=
{\bf q}_{\bot}^2/4m_e^2$, $\alpha=p'-p+1$, $\beta=p+p'+2m+1+1/B'$,
and
\begin{equation}\label{kfunction}
K^A_{lmpp'}({\bf q}_{\bot}^{\prime 2})=
\frac{2(-1)^{l+p+p^{\prime}}\, (2{\bf q}_{\bot}^{\prime 2})^l
\Gamma(l+m+1)}{ (l-p)!p!(l-p^{\prime})!p^{\prime}! \Gamma(m+1)}.
\end{equation}
Note that the various indices in the summation arise as follows:
the index $n$ comes from the photon-energy expansion, $l$ arises
from the binomial expansion of $((1-v^2)\omega^{\prime
2}+2(\cosh(zv)-\cosh(z)){\bf q}_{\bot}^{\prime 2}/z\sinh(z))^n$,
$p$ and $p'$ arises from writing $(\cosh(zv)-\cosh(z))^l$ as a sum
of exponential functions, and $m$ is due to expansions such as
$\sinh ^{-l} z = 2^l \exp [-lz] \sum^\infty_{m=0} C_m^{l+m-1}\exp
[-2mz]$. After performing the integrals in Eq.~(\ref{Aab}) and
replacing $n-l+1$ by $n$, we arrive at
\begin{eqnarray} \label{pipa}
\Pi_{\parallel}^A &=&-\frac{\alpha_e\omega^2}
{4\pi}\sin^2\theta\bar{\Pi}_{\parallel}^A= - \frac{ \alpha_e
m_e^2\omega^2}{\sqrt{\pi}q_{\parallel}^2} \sin^2\theta
{\sum}^{\prime} K^A_{lmpp'}({\bf q}_{\bot}^{\prime 2})
(B^{\prime})^{1-l} (\frac{q_{\parallel}^{\prime 2}}{\beta
B^{\prime}})^{n}\nonumber \\
&\times& \frac{\Gamma(n)} {\Gamma(n+\frac{1}{2})}\;
{}_2F_1(\frac{n}{2},\frac{n+1}{2};
n+\frac{1}{2};\frac{\alpha^2}{\beta^2}).
\end{eqnarray}
It is easy to calculate the derivatives of $\Pi_{\parallel}^A$.
Let us define
$F^A_{\parallel}(n,{\bf q}_{\bot}^2,B)=m_{\parallel}^{2n} {1\over
n!}{d^n\over d(q_{\parallel}^2)^n}
\Pi^A_{\parallel}{\vert}_{q_{\parallel}^2=0}$. We have
\begin{equation}\label{trans}
\kappa^A_{\parallel}= {1 \over 2i \omega} \int_C ds
F^A_{\parallel}(s,{\bf q}_{\bot}^2,B)u_{\parallel}^{-s},
\end{equation}
by inverting the sum rule in Eq.~(\ref{mellin2}). The above
integral transform is evaluated using
\begin{equation}\label{hyper}
\frac{1}{2\pi i}\int_C ds
x^{-s}\frac{\Gamma(s)}{\Gamma(s+\frac{1}{2})}
\;{}_2F_1(\frac{s}{2},\frac{s+1}{2};
s+\frac{1}{2};z^2)=\frac{1}{\sqrt{\pi}}
\frac{\Theta(1-x+\frac{x^2z^2}{4})} {\sqrt{1-x+\frac{x^2z^2}{4}}}.
\end{equation}
We arrive at
\begin{eqnarray}\label{abso2}
    \kappa^A_{\parallel}
&= &-  \frac{\alpha_e B'}{4q_{\parallel}^{\prime
2}}\omega\sin^2\theta {\sum}^{\prime \prime} {K^A_{lmpp'}({\bf
q}_{\bot}^{\prime 2}) \Theta\left((1-\frac{\beta
B^{\prime}}{q_{\parallel}^{\prime 2}}) +(\frac{\alpha
B^{\prime}}{2q_{\parallel}^{\prime 2}})^2\right)\over B^{\prime l}
\sqrt{(1-\frac{\beta B^{\prime}}{q_{\parallel}^{\prime 2}})
          +(\frac{\alpha B^{\prime}}
          {2q_{\parallel}^{\prime 2}})^2}},
\end{eqnarray}
where ${\sum}^{\prime \prime}\equiv \sum_{l=0}^\infty
\sum_{m=0}^{\infty}\sum_{p',p=0}^l$. To understand the structure
of $\kappa^A_{\parallel}$, we rewrite the denominator of the above
equation as
\begin{equation}\label{sing}
\sqrt{(1-\frac{\beta B^{\prime}}{q_{\parallel}^{\prime 2}})+
(\frac{\alpha B^{\prime}}{2q_{\parallel}^{\prime
2}})^2}=\frac{1}{q_{\parallel}^{\prime 2}}
\sqrt{q_{\parallel}^{\prime
4}-(\frac{1}{B'}+l_1+l_2)B^{\prime}q_{\parallel}^{\prime 2}
+\frac{1}{4}(l_1-l_2)^2 B^{\prime 2}},
\end{equation}
with $l_1=(\beta+\alpha-1/B')/2$ and $l_2=(\beta-\alpha-1/B')/2$.
It is easy to verify that this denominator has a root at
$q_{\parallel}^2=m_e^2(\sqrt{1+2l_1 B'}+\sqrt{1+2l_2 B'})^2$,
which is precisely the threshold value for $q_{\parallel}^2$ to
produce an electron-positron pair at the Landau levels $l_1$ and
$l_2$ respectively. This threshold behavior is also indicated by
the step function in the numerator.

Having worked out the $q_{\parallel}^2$ dependence of
$\kappa^A_{\parallel}$, we proceed to perform the summation in
$\Sigma^{\prime\prime}$. We first perform the summation over the
index $l$ in $K^A_{lmpp'}$:
\begin{eqnarray}
\sum_{l=0}^{\infty}\sum_{p,p'=0}^l\frac{
x^l\Gamma(l+m+1)}{(l-p)!(l-p')!} &=& \sum_{p,p'=0}^{\infty}
x^{\bar{p}}e^x\Gamma(\hat{p}+m+1)L_{\hat{p}+m}
^{\bar{p}-\hat{p}}(-x),
\end{eqnarray}
where $x=-2{\bf q}_{\bot}^{\prime 2}/B'\equiv -{\bf
q}_{\bot}^2/(2eB)$, $\bar{p}={\rm max} \{p,p'\}$, $\hat{p}={\rm
min}\{p,p'\}$, and $L_{\hat{p}+m} ^{\bar{p}-\hat{p}}$ is the
Laguerre polynomial. We then rewrite the left-over summation
$\sum_{p,p'=0}^{\infty}\sum_{m=0}^{\infty}$ as
$\sum_{l_1=1}^{\infty}\sum_{l_2=0}^{\infty}\sum_{m=0}^{\hat{l}}$
with $\hat{l}={\rm min}\{l_1-1,l_2\}$ using the relation
$l_1=(\beta+\alpha-1/B')/2=p'+m+1$ and $l_2=
(\beta-\alpha-1/B')/2=p+m$. The summation over the index $m$ can
now be performed as follows:
\begin{equation}
\sum_{m=0}^{\hat{l}}x^{-m}\frac{1}{\Gamma(\bar{l}-m+1)}
\frac{1}{\Gamma(\hat{l}-m+1)}\frac{1}{\Gamma(m+1)}=
\frac{x^{-\hat{l}}}{\Gamma(\bar{l}+1)}L_{\hat{l}}^{\bar{l}-\hat
{l}}(-x),
\end{equation}
where $\bar{l}={\rm max}\{l_1-1,l_2\}$. With the above summations
over $l$ and $m$, we arrive at
\begin{eqnarray}
    \kappa_{\parallel}^A\label{kaA}
&= &\frac{\alpha_e B'} {2q_{\parallel}^{\prime
2}}\omega\sin^2\theta
        \sum_{l_1=1}^{\infty}\sum_{l_2=0}^{\infty}
{{\bf T}^A_{l_1 l_2}(x) \Theta\left((1-\frac{\beta
B^{\prime}}{q_{\parallel}^{\prime 2}}) +(\frac{\alpha
B^{\prime}}{2q_{\parallel}^{\prime 2}})^2\right)\over
\sqrt{(1-\frac{\beta B^{\prime}}{q_{\parallel}^{\prime 2}})
          +(\frac{\alpha B^{\prime}}
          {2q_{\parallel}^{\prime 2}})^2}},
\end{eqnarray}
with\footnote{The absorption coefficient is not written in a
symmetrized form for saving the space. Nevertheless, the
symmetrization with respect to $l_1$ and $l_2$ can be easily done
as one wishes.}
\begin{equation}
{\bf T}^A_{l_1
l_2}(x)=(-1)^{1+r_A}x^{r_A}e^x\frac{\Gamma(\lambda_A+1)}
{\Gamma(\lambda_A+r_A+1)}L_{\lambda_A}^{r_A}(-x)L_{\lambda_A}^{r_A}(-x),
\end{equation}
where $r_A\equiv \bar{l}-\hat{l}=|l_1-l_2-1|$, and
$\lambda_A\equiv \hat{l}=(l_1+l_2-|l_1-l_2-1|-1)/2$.

Apart from the factor $\omega\sin^2\theta$, the absorption
coefficient $\kappa_{\parallel}^A$ is a function of $x\equiv -{\bf
q}_{\bot}^2/(2eB)$ and $q_{\parallel}^{\prime 2}$ where the latter
determines the threshold behaviors of the absorption coefficient.
Now that we have obtained the absorptive part of the polarization
function $\Pi_{\parallel}^A$, the dispersive part can be
calculated using the {\it Kramers-Kronig} relation. It is
convenient to recall the relation
$\Pi_{\parallel}^A=-(\alpha_e\omega^2 \sin^2\theta/ 4\pi)
\bar{\Pi}_{\parallel}^A(q_{\parallel}^{\prime 2},{\bf
q}_{\bot}^{\prime 2})$. Hence
\begin{equation}
{\rm Re}\bar{\Pi}_{\parallel}^A(q_{\parallel}^{\prime 2},{\bf
q}_{\bot}^{\prime 2})= \frac{P}{\pi}\int_{1}^{\infty}dt\frac{{\rm
Im} \bar{\Pi}_{\parallel}^A(t,{\bf q}_{\bot}^{\prime
2})}{t-q_{\parallel}^{\prime 2}},
\end{equation}
where $P$ stands for evaluating the principle part of the integral. We
point out that the refractive index for the polarization state,
$\epsilon^{\mu}_{\parallel}$, can be calculated through the
equation $q^2+{\rm Re}\Pi_{\parallel}=0$. Hence
\begin{equation}\label{refr}
1-n_{\parallel}^2(q_{\parallel}^{\prime 2},{\bf q}_{\bot}^{\prime
2},\theta)=\frac{\alpha_e\sin^2\theta}{4\pi^2}\;P
\int_{1}^{\infty}dt\frac{{\rm Im} \bar{\Pi}_{\parallel}(t,{\bf
q}_{\bot}^{\prime 2})}{t-q_{\parallel}^{\prime 2}}.
\end{equation}
It is important to note that the index of refraction
$n_{\parallel}$ also appears on the R.H.S. of the above equation,
through relations among the quantities, $q_{\parallel}^{\prime
2}$, ${\bf q}_{\bot}^{\prime 2}$, and the angle $\theta$. Thus the
calculation of $n_{\parallel}$ is in principle nontrivial even if
the polarization function ${\rm Re}\Pi_{\parallel}$ is known. Let
us consider a simplified case where the magnetic-field strength is
super-critical, i.e., $B\gg B_c$. In this case, the quantity
$1-n_{\parallel}^2$ to the leading order in $B_c/B$ is given by
the lowest Landau-level contribution to the dispersion integral in Eq.~(\ref{refr}).
Hence, we arrive at\cite{KLT2}
\begin{equation}\label{super}
n_{\parallel}^2(q_{\parallel}^{\prime 2}, {\bf q}_{\bot}^{\prime 2}, \theta)-1=\frac{\alpha_e
B'\sin^2\theta}{2\pi q_{\parallel}^{\prime 2}}
\left(\frac{1}{\sqrt{q_{\parallel}^{\prime
2}(1-q_{\parallel}^{\prime 2})}}
\arctan\sqrt{\frac{q_{\parallel}^{\prime
2}}{1-q_{\parallel}^{\prime 2}}}
-1\right)\exp(-\frac{2{\bf q}_{\bot}^{\prime 2}}{B^{\prime}}).
\end{equation}
It is instructive to rewrite $\exp(-2{\bf q}_{\bot}^{\prime
2}/B^{\prime})$ as $\exp(-2\omega^{\prime 2}\sin^2\theta
n_{\parallel}^2/B^{\prime})$ with $\omega^{\prime}=\omega/2m_e$.
Clearly, for $q_{\parallel}^2$ not very close to $4m_e^2$ (the
threshold energy for the pair production), one can safely set
$\exp(-2{\bf q}_{\bot}^{\prime 2}/B^{\prime})=1$. The index of
refraction $n_{\parallel}^2$ with this approximation then agrees
with the previous result\cite{ckm}. However, as
$q_{\parallel}^{\prime 2}$ is so close to 1 such that
$\alpha_e/\sqrt{1-q_{\parallel}^{\prime 2}}$ becomes greater than
unity, one can no longer neglects the exponential factor
$\exp(-2{\bf q}_{\bot}^{\prime 2}/B^{\prime})$. In fact, in the
limit $q_{\parallel}^{\prime 2}\to 1$, one finds that ${\bf
q}_{\bot}^{\prime 2}\to \infty$, as pointed in the previous
literature\cite{shabad,usov}. We note that the behaviors of
another polarization state, $\epsilon_{\bot}^{\mu}$, can be
studied in the similar manner. The details are given in
Ref.~\cite{KLT2}.

For consistency, it is desirable to compare our absorption
coefficients with those obtained by squaring the $\gamma\to
e^+e^-$ amplitude directly\footnote{We make comparisons with the
most updated results in Ref. \cite{DH}.}. Our results reduce to
that of Daugherty and Harding in the special limit $q^2=0$, which
is an assumption made by these authors in their calculations. Our
finding is similar to what Shabad has demonstrated in Ref.
\cite{prep} as he compared his result with that of Klepikov
\cite{kle} in the same limit for $q^2=0$. Our approach
differs from that of Refs. \cite{shabad,prep} in that Shabad
performs the calculation in the beyond-threshold energy where the
algebraic manipulations are rather involved and precautions are
required, whereas we take the advantage of inverse Mellin
transform which permits us to calculate the polarization function
near the zero longitudinal momentum with a convenient momentum
expansion.

\section{Conclusion}

In conclusion, we have developed an integral-transform
technique to compute the absorptive part of photon polarization
function in a background magnetic field, while the dispersive part
can be obtained via the {\it Kramers-Kronig} relation. We
emphasize that this technique is rather powerful for the current
problem in which there are infinite numbers of singularity
occurring in the real positive axis of $q_{\parallel}^{\prime 2}$,
each corresponds to an $e^{+}-e^{-}$ system in a specific combination of
Landau levels. This technique has many other applications. To name
a few, one may analyze the photon polarization function in a
general background electromagnetic field, or study other
current-current correlation functions under the same external
condition. To our knowledge, the vector-axial vector correlation
function relevant to weak interaction processes has not been
analyzed as detailed as the fashion presented in this work. We
shall report the results of such analysis as well as other
relevant studies in a forthcoming publication\cite{kl}.

\acknowledgements
I wish to thank Prof. M. Bordag for organizing this workshop. This
work is supported by Taiwan's National Science Council under the
grant numbers: NSC89-2112-M009-041 and NSC90-2112-M-009-023.

\eject
\end{document}